\documentclass[10pt]{article}
\usepackage{amsmath, latexsym, amssymb, amscd, amsfonts, amsthm}
\usepackage[english]{babel}

\DeclareFixedFont{\sfracFont}{U}{euf}{b}{n}{7pt}
\newcommand{\BB}[1]{\mathbb{#1}}
\newcommand{\p}{|\!|}

\newcommand{\I}{\mathbf{1}}

\begin{document}
\title{Bellman equations for optimal feedback control of qubit states}
\author{Luc Bouten, Simon Edwards and V.P.~Belavkin}
\date{}
\maketitle

\begin{abstract}
Using results from quantum filtering theory
and methods from classical control theory,
we derive an optimal control strategy for
an open two-level system (a qubit in
interaction with the electromagnetic field)
controlled by a laser. The aim is to optimally
choose the laser's amplitude
and phase in order to drive the system
into a desired state. The Bellman equations
are obtained for the case of diffusive and
counting measurements for vacuum field states.
A full exact solution of
the optimal control problem is given
for a system with simpler, linear, dynamics.
These linear dynamics can be obtained
physically by considering a two-level 
atom in a strongly driven, heavily damped, 
optical cavity.
\end{abstract}

\section{Introduction}

The advent of quantum information theory and
the ever increasing experimental possibilities
to implement this theory on real physical
systems e.g.\ \cite{AASDM02}, \cite{GSDM03},
has created great demand for a theory
on the control of
quantum systems. Since qubits, i.e.\ two-level
quantum systems, make up the hardware for
quantum information processing, one important
question is how to optimally control or
engineer their states. Many problems of quantum
computation can be formulated in terms of quantum
optimal control of unitary or decohering gates.
Most previous work on the optimal control of qubit states
use an open loop strategy with a variational calculus
approach to optimization \cite{PK02}, \cite{XYOFR04},
\cite{TV02}.  However, in order to apply controls one
must consider the qubit as an open quantum system
which gives the possibility for time-continuous non
demolition measurements and thus a closed (feedback)
loop strategy would be more advantageous.  In this
paper, we employ a feedback strategy using dynamic
programming which is a globally optimal solution to
the control problem and thus extends the previous
locally optimal variational approaches.

The importance of feedback control theory in
the control of open quantum systems was first
recognized by Belavkin in \cite{Bel83}.
Like in the classical case with partially
observed systems, a feedback control
strategy is usually favorable to the open
loop control (without feedback).
Optimal feedback control strategies for the
open quantum
oscillator appeared even earlier in \cite{Bel79} and a
quantum Bellman equation for optimal feedback
control was introduced in \cite{Bel88} for a
general diffusive and a counting measurement process.
An interest in optimal quantum control and stability
theory has recently emerged in the optics community
\cite{DHJMT00}, \cite{HSM04}, \cite{Kol92}.

As it was shown in the above papers, since we never
have complete observability of
quantum systems, the problem of quantum
feedback control must involve a filtering procedure
in order to measure and control the system
optimally. We can separate these two problems as was
suggested in \cite{Bel83} and consider first the
problem of quantum filtering \cite{Bel80},
\cite{Bel88}, \cite{Bel92a}, \cite{Bel92b},
\cite{BGM04}. In quantum filtering theory pioneered by
Belavkin in \cite{Bel79}-\cite{Bel88}, the quantum filtering
equation for the system with a chosen continuous
non demolition
measurement has to be derived. A system observed
through its interaction with the electromagnetic
field by continuous measurement of some field
observables, needs to be updated continuously in time
to incorporate the information gained by the
measurement. That is we have to condition the quantum
state of the system on the obtained measurement
results continuously in time. The quantum filtering
equation as it was first introduced in \cite{Bel80},
\cite{Bel88} is a stochastic differential equation
for the conditioned state in which the innovation
process, representing the information gain, is one of
the driving terms. Like in the quantum optics literature,
we take the filtering equation as our starting point,
however, the driving Wiener process is not treated as
the noise, but as an innovation process.  For more
background on the derivation of this stochastic
equation as a general filtering equation in an open
quantum system conditioned with respect to a non
demolition observation, see \cite{Bel92a},
\cite{Bel92b}, \cite{BGM04}.

Once the quantum filtering
equation is obtained, we are left with a
classical control problem.
In particular, if the state of a qubit is
parameterized by
its polarization vector in the Bloch sphere,
i.e.\ a vector in
the $3$-dimensional unit ball providing
sufficient coordinates for the system \cite{Bel83},
the filtering equation provides
stochastic dynamics for the polarization
vector. The control is present in the dynamics
through Rabi oscillations, which perform
rotations of the polarization vector in the
Bloch sphere
caused by a laser driving the qubit.
The phase and intensity of the laser
are the parameters that can be controlled.

The main aim of this paper is to demonstrate the
relevance of classical control and quantum
filtering when controlling
quantum systems.  This is shown by the
example of optimal control of a
two-level quantum system. A cost function,
which is a measure of optimality of the control,
is introduced and the corresponding Bellman
equations are derived for this system.  From
these equations, we produce an optimal
control strategy which depends on the
solutions to the corresponding
Hamilton-Jacobi-Bellman equation.
In general these solutions are very difficult
to find, even numerically, so we resort to a
physically motivated simplification of the
dynamics by considering a qubit in strongly 
driven, heavily damped, optical cavity.
This enables us to present an exact solution to
the control problem.

The remainder of the paper is organized
into four main sections.  Firstly we describe
the model and introduce the dynamics of the
polarization vector from the filtering equation
for diffusive and counting measurement for an initial 
vacuum field state. 
The next section describes
the \emph{principle of optimality} which is the
key idea behind optimal feedback control and
enables us to derive the Bellman equations in
Section 4.  We finish the paper with the
simpler model corresponding to a two-level 
system in a strongly driven, heavily damped, 
optical cavity. Here we obtain a linear
filtering equation, which we use with a quadratic
cost function to give an exact solution for
the optimal feedback control strategy.

\section{The model and state dynamics}

We consider a two-level system, i.e.\ a qubit,
in interaction with the quantized electromagnetic field
in the weak coupling limit \cite{Dav74}, \cite{AFLu90}.
This means that the unitary dynamics of the qubit 
and the field together in the interaction picture
is given by a quantum stochastic differential equation
(QSDE). In this way the field acts as non-commutative
noise on the qubit.
The initial state of the noise (electromagnetic field) is
taken to be the vacuum state and the reduced dynamics
of the qubit is given by a master equation.  Such a
quantum Langevin model was the starting point of the
quantum stochastic theory of continuous non demolition
measurements developed in \cite{Bel80}-\cite{Bel92a}.
 
We control the state of the qubit by its interaction with
a laser beam. This laser beam is given by a channel
in the field, called the \emph{forward channel}, which
is in a coherent state $\psi(u)$, where $u$ is a square
integrable complex valued function of time.
The control function $u$ induces Rabi oscillations which we
must choose carefully to rotate the state of the qubit in the
desired manner. The rest of the field
is called the \emph{side channel}. We assume that there is
no direct scattering between the two channels.
Following \cite{Bel88} and \cite{Bel92a}, we consider two
different continuous time measurement schemes to be
performed in the side channel. The first measurement
scheme we consider is a homodyne detection experiment
which measures the field quadrature $Y_t = A^*_s(t) +
A_s(t)$ which is a classical diffusive observable
process at the output of the quantum system. The
second scheme is a counting experiment, counting the
number $Y_t=N_t$ of fluorescence photons emitted by the
qubit up to time $t$.

Since the side channel and atom are in interaction, we gain
information on how the state of the qubit changes
from the measurement results of the homodyne detection
experiment or the counting experiment. The state of the qubit
conditioned on the measurement result $\omega$ of the
non demolition output process ${Y_t}$ is a
\emph{random} state. This means it is a map
$\rho^t_\bullet$ from the possible paths of
measurement results $\Omega$ to the $2 \times
2$-density matrices, mapping $\omega \in \Omega$ to
the density matrix $\rho^t_\omega$ which represents
the state of the qubit conditioned on a path of
measurement results $\omega$ up to time $t$. Note
that for homodyne detection, a path $\omega$ of
measurement results is just the path of the
photocurrent from time $0$ to time $t$. For the
counting experiment a path of measurement results is
given by the collection of times at which photons
were detected.

The conditional state
evolution of the qubit is given by a
classical stochastic differential equation for
the density matrix $\rho^t_\bullet$ called the quantum
filtering or Belavkin equation \cite{Bel88},\
\cite{Bel92b},\ \cite{BGM04}. For the homodyne
detection experiment we take the quantum filtering
qubit equation derived in \cite{Bel92b} with respect to
the diffusive output process $Y_t$, as our starting
point
  \begin{equation}\label{eq Belavkin}\begin{split}
  d\rho^t_\bullet = L(&\rho^t_\bullet)dt\ + \
  \Big(V_s\rho^t_\bullet + \rho^t_\bullet V_s^* -
  \mbox{Tr}\big(V_s\rho^t_\bullet + \rho^t_\bullet V_s^*\big)\rho^t_\bullet\Big)\ \times \\
  &\Big(dY_t -\mbox{Tr}\big(V_s\rho^t_\bullet + \rho^t_\bullet V_s^*\big)dt\Big)
  \end{split}\end{equation}
where
  \begin{equation}\label{eq V}
  V_s:= \kappa_sV\ \ \ \ \mbox{with}\ \ \ \ V:= \begin{pmatrix} 0 & 0 \\ 1 & 0 \end{pmatrix},
  \end{equation}
and $\kappa_s^2$ is the decay rate into the side channel. Furthermore,
the Lindblad term $L$ is given by
  \begin{equation}\label{eq Lindblad}
  L(\rho) =  -i[H,\rho] + V_f\rho V_f^* - \frac{1}{2}\{V_f^*V_f, \rho\}
  + V_s\rho V_s^* -\frac{1}{2}\{V_s^*V_s, \rho\},
  \end{equation}
with time dependent controlling Hamiltonian
  \begin{equation*}
  H := \begin{pmatrix}0 & -i\kappa_f u(t) \\ i\kappa_f\overline{u}(t) & 0\end{pmatrix},
  \end{equation*}
and with $V_f := \kappa_fV$ where $\kappa_f^2$ is the decay rate into the forward channel. We
choose units such that $\kappa_s^2 + \kappa_f^2 = 1$. The form of the
Hamiltonian physically relates to two orthogonal control fields
corresponding to the real and imaginary parts of the complex control
function $u(t)$.  The innovating martingale (second line of \eqref{eq
Belavkin}) is just a Wiener process $W_t$ which describes the information
gain from the measurement, i.e.\ the observed process $Y_t$ satisfies the
stochastic differential equation
  \begin{equation}\label{eq obsproc}
  dY_t :=  \mbox{Tr}\big(V_s\rho^t_\bullet +
  \rho^t_\bullet V^*_s\big)dt + dW_t,
  \end{equation}
and the Belavkin diffusion filtering equation
\eqref{eq Belavkin} can be written as a stochastic
master equation
  \begin{equation}\label{eq qfilt}
  d\rho^t_\bullet = L(\rho^t_\bullet)dt +
  \Big(V_s\rho^t_\bullet + \rho^t_\bullet V^*_s -
  \mbox{Tr}\big(V_s\rho^t_\bullet + \rho^t_\bullet V^*_s\big)\rho^t_\bullet\Big)dW_t.
  \end{equation}

For ${\bf a} \in \BB{R}^3$ we introduce the notation
$\sigma({\bf a}) := a_1\sigma_x + a_2 \sigma_y + a_3\sigma_z$, where
$\sigma_x, \sigma_y$ and $\sigma_z$ denote the Pauli spin matrices.
The states of a qubit can be parameterized by vectors in
the Bloch sphere $B := \{{\bf p} \in \BB{R}^3;\ \p {\bf p} \p \le 1\}$.
The random vector with which we parameterize the state
$\rho^t_\bullet$ is denoted ${\bf P}^t$ and is called its
\emph{polarization vector}, i.e.\ we write
  \begin{equation*}
  {\bf P}^t = \begin{pmatrix} P^t_x \\ P^t_y \\ P^t_z
   \end{pmatrix}, \ \ \ \ \ \ \ \ \ \   
  \rho^t_\bullet = \frac{\I
  +\sigma({\bf P}^t)}{2}.   
  \end{equation*} Introducing $u^+_t :=
\kappa_f\mbox{Re}(u(t))$ and $u^-_t := \kappa_f\mbox{Im}(u(t))$ we can write 
the filtering equation \eqref{eq qfilt} as
  \begin{equation}\label{eq Belvac}
    d{\bf P}^t = \begin{pmatrix}
  -\frac{1}{2}P^t_x - 2u^+_tP^t_z \\
  -\frac{1}{2}P^t_y + 2u^-_tP^t_z \\
  -(1 + P^t_z)
  +2u^+_tP^t_x - 2u^-_tP^t_y
  \end{pmatrix}dt\ +
  \ \begin{pmatrix}1 + P^t_z - {P^t_x}^2  \\
  -P^t_xP^t_y \\
  -P^t_x(1+P^t_z)
  \end{pmatrix}\kappa_sdW_t.
  \end{equation}

For the counting experiment, the Belavkin filtering
equation derived in \cite{Bel88} reads as
  \begin{equation*}
  d\rho^t_\bullet = L(\rho^t_\bullet)dt +
  \Big(\frac{V_s\rho^t_\bullet V_s^*}{\mbox{Tr}(V_s\rho^t_\bullet V_s^*)}- \rho^t_\bullet\Big)
  \Big(dN_t - \mbox{Tr}(V_s\rho^t_\bullet V_s^*)dt\Big),
  \end{equation*}
where $L$ and $V_s$
are given by \eqref{eq Lindblad} and \eqref{eq V},
and $N_t$ is the random variable counting the number
of detected photons up to time $t$. In parameterized
form this reads
  \begin{equation}\label{eq Belcount}
     d{\bf P}^t = \begin{pmatrix}
  -\frac{1}{2}P^t_x - 2u^+_tP^t_z \\
  -\frac{1}{2}P^t_y + 2u^-_tP^t_z \\
  -\big(1 + P^t_z\big)
  +2u^+_tP^t_x - 2u^-_tP^t_y
  \end{pmatrix}dt +
  \begin{pmatrix}
  - P^t_x \\
  - P^t_y \\
  -\big(1+ P^t_z\big)
  \end{pmatrix}\Big(dN_t - \frac{\kappa_s^2}{2}\big(1+P^t_z\big)dt\Big).
  \end{equation}

\section{The principle of optimality}

In order to find an optimal quantum feedback control strategy based on the
continuous non demolition observation we shall use the dynamic programming
method for the sufficient qubit coordinate ${\bf P}^t$ as it was
suggested in \cite{Bel83}.

At time $t=0$ the qubit is taken to be
in a known initial state ${\bf P}^0$. It is our
objective to bring it in the $\sigma_z$-up
state at time $t=T$, at which the
control experiment is stopped. This is
done by choosing the
laser intensity and phase, given in terms
of $u_t^+$ and $u_t^-$, at every time $t$
which may depend on the stochastic
state ${\bf P}^t$ of the qubit at time $t$,
via a feedback mechanism.
The total cost of the control experiment
from time $0$ up to time $T$ is described by
  \begin{equation}\label{eq J}
  J := \big(1-P^T_z\big) + \int_0^T \big({u^+_s}^2 + {u^-_s}^2\big) ds.
  \end{equation}
The first term reflects our main objective which is to bring the
system in the $\sigma_z$-up state at time $T$, whereas
the second term reflects the cost for using the
laser. The second term restricts our resources.
Without this restriction it would be possible to
apply brute force, e.g.\ a very strong laser pulse
at the end of the experiment, to obtain our goal.

Note that the total cost $J$ of equation \eqref{eq J} is
a random variable. It depends on the stochastic measurement
results through the random variable $P^T_z$ and the
applied controls $u_t^+$ and $u_t^-$, which in their turn
depend on the random state ${\bf P}^t$ of the
qubit. From equation $\eqref{eq J}$ it follows
that the \emph{expected cost-to-go} $J(t,{\bf P}^t)$
at time $t$ when we are in the state ${\bf P}^t$
at time $t$, is given by
  \begin{equation}\label{eq cost}
  J(t, {\bf P}^t) := \BB{E}_{{\bf P}^t}\Big[
  \big(1-P^T_z\big) + \int_t^T \big({u^+_s}^2 + {u^-_s}^2\big) ds\Big],
  \end{equation}
where $\BB{E}_{{\bf P}^t}$ denotes the expectation over
all possible paths of measurement results conditioned
on the fact that we are in state ${\bf P}^t$ at time
$t$. The problem addressed in this paper is how
to choose the feedback controls $u^+_t$ and $u^-_t$
at every time $t$, such
that the total expected cost $J(0,{\bf P}^0) \ (=\BB{E}_{{\bf P}^0}[J])$ is
\emph{minimal}. The solution to this
problem, i.e.\ a map $\mu^*:\ [0,T]\times B \to \BB{R}^2$
assigning numbers $u^+_t$ and $u^-_t$ to every time $t$
and state ${\bf P}^t$ that
minimize $J(0, {\bf P}^0)$, is called an
\emph{optimal strategy}. Here the star $^*$ in
$\mu^*$ denotes optimality of the strategy. Extending
this convention we denote the minimal or
optimal cost by $J^*(0,{\bf P}^0)$.

A key observation in this problem is that if we have
a strategy $\mu_{[s,T]}^*$, $0<s<T$ that is optimal over the interval $[s,T]$
(i.e. one which minimizes $J(s, {\bf P}^s)$ for every possible state
${\bf P}^s$ at time $s$) then the optimal strategy $\mu^*$
of the whole experiment coincides with $\mu_{[s,T]}^*$ when restricted
to the interval $[s,T]$.  So we optimize over disjoint intervals,
working backwards in time to build an optimal strategy for the whole
experiment.  This idea is called the \emph{principle of
optimality} \cite{Bel57} and lies at the heart of
optimal feedback control theory.

Iteration of the principle of optimality enables
a recursive solution to the optimal control
problem known as \emph{dynamic programming}
\cite{Bel57}. To illustrate this method
we divide the time interval $[0,T]$ into
$N$ parts of equal size $\Delta := T/N$.
The principle of optimality leads for $0 \le n <N$ to
the following recursive \emph{dynamic programming equation}
\cite{Bel57}, \cite{Kus71}
 \begin{equation}\label{eq DP}
  J^*(n,\, {\bf P}^n) =  \min_{u^+_{n}, u^-_{n}}
  \Big\{\BB{E}_{{\bf P}^n}\Big(\big({u^+_{n}}^2 + {u^-_{n}}^2\big)\Delta +
  J^*(n+1,\, {\bf P}^{n+1})\Big)\Big\},
  \end{equation}
with boundary condition $J^*(N,\, {\bf P}^N) = 1 - P^N_z$.
Using the state dynamics, ${\bf P}^{n+1}$ can be expressed
in terms of ${\bf P}^{n}$, $u^+_n$ and $u^-_n$.
The minimization of \eqref{eq DP} working backwards from $n=N-1$
to $n=0$ yields the optimal control strategy $(u^+_n,u^-_n)=\mu^*(n,{\bf P}^n)$.

In the next section we derive a partial differential
equation for the expected optimal cost to go $J^*$
by studying equation \eqref{eq DP} with boundary
condition $J^*(T,\, {\bf P}^T) = 1 - P^T_z$ in
infinitesimal form
  \begin{equation}\label{eq inf}
  J^*(t,\, {\bf P}^t) =  \min_{u^+_{t}, u^-_{t}}
  \Big\{\BB{E}_{{\bf P}^t}\Big(\big({u^+_{t}}^2 + {u^-_{t}}^2\big)dt +
  J^*(t+dt,\, {\bf P}^{t+dt})\Big)\Big\}.
  \end{equation}
This is done by using the state dynamics for
${\bf P}^{t+dt}$ and by subsequently expanding
$J^*$ up to the correct order according to
It\^o's formula. Solving the obtained partial differential
equation is equivalent to
running the dynamic programming algorithm
and therefore provides a solution to the optimal
control problem.

\section{Bellman equations}

In this section we first consider the case where
we are measuring the field quadrature $Y_t = A^*_s(t)+A_s(t)$
by a homodyne detection scheme. The dynamics are
given by equation \eqref{eq Belvac}.
According to It\^o's formula we have
  \begin{equation}\label{eq Ito}\begin{split}
  dJ^*(t,{\bf P}^t)
  ={}&
  \partial_t J^*(t,{\bf P}^t)dt + \sum_{\mu = x,y,z}
  \partial_\mu J^*(t,{\bf P}^t)dP^t_\mu\ + \\  
  &\frac{1}{2}
  \sum_{\mu,\nu = x,y,z} \partial^2_{\mu\nu} J^*(t,{\bf P}^t)dP^t_\mu dP^t_\nu,
  \end{split}\end{equation}
where $\partial_\mu$ denotes partial differentiation
of $J^*(t,{\bf P}^t)$ with respect to $P^t_\mu$ and $\partial^2_{\mu\nu}$
denotes partial differentiations with respect to $P^t_{\mu}$ and $P^t_{\nu}$
where we assume that $J^*$ is suitably differentiable. Using the state
dynamics \eqref{eq Belvac}, the differentials $dP^t_\mu$ can be expressed
in terms of $dt$ and $dW_t$ and products of differentials can be evaluated
using It\^o's rule $dW_tdW_t =dt$. Since the expectation of the innovating
martingale is zero, i.e.\ $\BB{E}_{{\bf P}^t}[dW_t] = 0$, equation
\eqref{eq inf} simplifies a great deal by substituting $J^*(t+dt, {\bf
P}^{t+dt}) = J^*(t, {\bf P}^t) + dJ^*(t, {\bf P}^t)$ and using \eqref{eq
Ito} to obtain
  \begin{equation}\label{eq Bellvac}\begin{split}
   &-\partial_t J^* =  \min_{u^+_{t}, u^-_{t}}
  \Bigg\{
  {u^+_{t}}^2 + {u^-_{t}}^2
  - 2u^+_tP^t_z\partial_x J^*
  + 2u^-_tP^t_z\partial_y J^*
  +\big(2u^+_tP^t_x - 2u^-_tP^t_y\big)\partial_z J^*
  \Bigg\}\ +\\
  &\kappa_s^2\Bigg(
  \big({P^t_x}^2-1-P^t_z\big)P^t_xP^t_y\partial^2_{xy} J^* +
  \big({P^t_x}^2-1-P^t_z\big)P^t_x\big(1+P^t_z\big)\partial^2 _{xy}J^* +
  {P^t_x}^2P^t_y\big(1+P^t_z\big)\partial^2_{yz} J^*
  \Bigg)\ + \\
  &\frac{\kappa_s^2}{2}\Bigg(
  \Big(1 + P^t_z - {P^t_x}^2\Big)^2\partial^2_{xx} J^* +
  {P^t_x}^2{P^t_y}^2\partial^2_{yy} J^* +
  {P^t_x}^2\big(1+P^t_z\big)^2\partial^2_{zz} J^*
  \Bigg)\ - \\
  &\Bigg(\frac{1}{2}P^t_x\partial_x J^*
  +\frac{1}{2}P^t_y\partial_y J^*
  +\big(1 + P^t_z\big)\partial_z J^*\Bigg),
  \end{split}\end{equation}
with boundary condition $J^*(T,\, {\bf P}^T) = 1 - P^T_z$.
In control theory, equation \eqref{eq Bellvac} is
known as the \emph{Bellman equation} and it was introduced into quantum
feedback control theory in \cite{Bel88}.

We evaluate the minimum in the Bellman equation \eqref{eq Bellvac}
by completing the squares on $u^+_t$ and
$u^-_t$
  \begin{equation*}\begin{split}
  &{u^+_{t}}^2 + {u^-_{t}}^2
  - 2u^+_tP^t_z\partial_x J^*
  + 2u^-_tP^t_z\partial_yJ^*
  +\big(2u^+_tP^t_x - 2u^-_tP^t_y\big)\partial_zJ^* = \\
  &\Big(u^+_t + \big(P^t_x\partial_zJ^* -
  P^t_z\partial_xJ^*\big)\Big)^2 -
  \big(P^t_x\partial_zJ^* - P^t_z\partial_xJ^*\big)^2\ + \\
  &\Big(u^-_t + \big(P^t_z\partial_yJ^* -
  P^t_y\partial_zJ^*\big)\Big)^2 -
  \big(P^t_z\partial_yJ^* - P^t_y\partial_zJ^*\big)^2.
  \end{split}\end{equation*}
In this way we find an optimal control strategy in terms of the
partial derivatives of the optimal expected cost-to-go, given
by
  \begin{equation}\label{eq strategy}
   u^+_t = P^t_z\partial_xJ^* - P^t_x\partial_zJ^*,\ \ \ \
   u^-_t = P^t_y\partial_zJ^* - P^t_z\partial_yJ^*.
  \end{equation}
where the optimal expected cost-to-go $J^*$
is the solution to the following second order non-linear
partial differential equation
  \begin{equation}\label{eq part}\begin{split}
  &-\partial_tJ^* = \\
  &\kappa_s^2\Bigg(
  \big({P^t_x}^2-1-P^t_z\big)P^t_xP^t_y\partial^2_{xy}J^* +
  \big({P^t_x}^2-1-P^t_z\big)P^t_x\big(1+P^t_z\big)\partial^2_{xy}J^* +
  {P^t_x}^2P^t_y\big(1+P^t_z\big)\partial^2_{yz}J^*
  \Bigg)\ + \\
  &\frac{\kappa_s^2}{2}\Bigg(
  \Big(1 + P^t_z - {P^t_x}^2\Big)^2\partial^2_{xx}J^* +
  {P^t_x}^2{P^t_y}^2\partial^2_{yy}J^* +
  {P^t_x}^2\big(1+P^t_z\big)^2\partial^2_{zz}J^*
  \Bigg)\ - \\
  &\Bigg(\frac{1}{2}P^t_x\partial_xJ^*
  +\frac{1}{2}P^t_y\partial_yJ^*
  +\big(1 + P^t_z\big)\partial_zJ^*\Bigg)
  - \big(P^t_x\partial_zJ^* - P^t_z\partial_xJ^*\big)^2
  - \big(P^t_z\partial_yJ^* - P^t_y\partial_zJ^*\big)^2,
  \end{split}\end{equation}
with boundary condition $J^*(T,\, {\bf P}^T) = 1 - P^T_z$.
This type of equation is called a \emph{Hamilton-Jacobi-Bellman} (HJB) equation.
However, even finding a numerical solution to this equation is still a very hard
problem which is beyond the scope of this paper.
In the following section we will look at a system with
much simpler dynamics for which we can actually solve the
HJB-equation.

In the remainder of this section
we turn our attention to
the situation where we count photons
in the side channel.
We consider the same problem as before, i.e.\
we want to find optimal controls $u^+_t$
and $u^-_t$ depending on ${\bf P}^t$ for each time $t$, such
that the total expected cost $J(0,\, {\bf P}^0)$ of equation
\eqref{eq cost} is minimal. Since $N_t$ is a
jump process, we use the It\^o rule $dN_tdN_t = dN_t$ and the It\^o formula
for calculating $dJ^*(t,\, {\bf P}^t)$ has
also changed. Using the dynamics \eqref{eq Belcount} we find
 \begin{equation}\label{eq If}\begin{split}
  dJ^*(t,{\bf P}^t) ={} &\partial_tJ^*(t,{\bf P}^t)dt +
  \Big(\frac{\kappa_s^2}{2}P^t_x\big(1+P^t_z\big)
  -\frac{1}{2}P^t_x - 2u^+_tP^t_z \Big)\partial_xJ^*(t,{\bf P}^t)dt\ + \\
  &\Big(\frac{\kappa_s^2}{2}P^t_y\big(1+P^t_z\big)
  -\frac{1}{2}P^t_y + 2u^-_tP^t_z\Big)J^*_y(t,{\bf P}^t)dt\ + \\
  &\Big(\frac{\kappa_s^2}{2}\big(1+P^t_z\big)^2
  -\big(1 + P^t_z\big)+2u^+_tP^t_x - 2u^-_tP^t_y\Big)
  \partial_zJ^*(t,{\bf P}^t)dt\ + \\
  & \Big(J^*(t, {\bf P}^t + {\bf Q}^t)-J^*(t, {\bf P}^t)\Big)dN_t,
  \end{split}\end{equation}
where ${\bf Q}^t$ in the difference term is given by
  \begin{equation*}
    {\bf Q}^t :=\begin{pmatrix}
  - P^t_x \\
  - P^t_y \\
  -\big(1+ P^t_z\big)
  \end{pmatrix},\ \ \ \mbox{i.e.}\ \ \ \
  {\bf P}^t + {\bf Q}^t =
  \begin{pmatrix}0 \\ 0 \\ -1\end{pmatrix}.
  \end{equation*}

Starting from the equation \eqref{eq inf}, using
$J^*(t+dt,\, {\bf P}^{t+dt}) = J^*(t,\, {\bf P}^t) + dJ^*(t,\, {\bf P}^t)$,
It\^o's formula \eqref{eq If} and the fact that
$\BB{E}_{{\bf P}^t}\big[dN_t\big] =
\frac{\kappa_s^2}{2}\big(1+P^t_z\big)dt$, we find
the following Bellman equation for the photon
counting case (cf.\ \cite{Kol92})
  \begin{equation}\begin{split}\label{eq Bellcount}
  -\partial_tJ^* = {}
  &\min_{u^+_{t}, u^-_{t}}
  \Bigg\{{u^+_{t}}^2 + {u^-_{t}}^2
  - 2u^+_tP^t_z \partial_xJ^* + 2u^-_tP^t_z\partial_yJ^* +
  \big(2u^+_tP^t_x - 2u^-_tP^t_y\big)\partial_zJ^*\Bigg\}\ + \\
  & \frac{\kappa_s^2}{2}\big(1+P^t_z\big)
  \Big(J^*(t, {\bf P}^t + {\bf Q}^t)-J^*(t, {\bf P}^t)\Big) +
  \Big(\frac{\kappa_s^2}{2}P^t_x\big(1+P^t_z\big)
  -\frac{1}{2}P^t_x\Big)\partial_xJ^* \ + \\
  &\Big(\frac{\kappa_s^2}{2}P^t_y\big(1+P^t_z\big)
  -\frac{1}{2}P^t_y \Big)\partial_yJ^* +
  \Big(\frac{\kappa_s^2}{2}\big(1+P^t_z\big)^2
  -\big(1 + P^t_z\big)\Big)\partial_zJ^*,
  \end{split}\end{equation}
where the partial derivatives are all
evaluated at $(t, {\bf P}^t)$ and the boundary
condition is
$J^*(T,\, {\bf P}^T) = 1 - P^T_z$. Completing
the squares leads again to an optimal
control strategy given by equation \eqref{eq strategy},
where $J^*$ in the case of photon counting has to satisfy
the following HJB equation
  \begin{equation*}\begin{split}
  -\partial_tJ^* = {}
     & \frac{\kappa_s^2}{2}\big(1+P^t_z\big)
  \Big(J^*(t,\, {\bf P}^t + {\bf Q}^t)-J^*(t,\, {\bf P}^t)\Big) +
  \Big(\frac{\kappa_s^2}{2}P^t_x\big(1+P^t_z\big)
  -\frac{1}{2}P^t_x\Big)\partial_xJ^* \ + \\
  &\Big(\frac{\kappa_s^2}{2}P^t_y\big(1+P^t_z\big)
  -\frac{1}{2}P^t_y \Big)\partial_yJ^* +
  \Big(\frac{\kappa_s^2}{2}\big(1+P^t_z\big)^2
  -\big(1 + P^t_z\big)\Big)\partial_zJ^* \ - \\
  &\big(P^t_x\partial_zJ^* - P^t_z\partial_xJ^*\big)^2 -
  \big(P^t_z\partial_yJ^* - P^t_y\partial_zJ^*\big)^2,
  \end{split}\end{equation*}
with boundary condition $J^*(T,\, {\bf P}^T) = 1 - P^T_z$.
Solving this equation is again beyond the scope of
this paper.

\section{A simpler model}

As we have discovered in the previous section,
realistic optimal control problems usually
lead to very difficult Bellman equations. In this section
we will study a drastically more simple model with 
linear dynamics given by
  \begin{equation}\label{eq Belcom}
  d\rho^t_\bullet = L(\rho^t_\bullet)dt - i\alpha[\sigma_z, \rho^t_\bullet]dW_t,
  \end{equation}
with $\alpha$ a real constant and
  \begin{equation*}
  L(\rho) =  -i[B_t\sigma_z,\rho] + \alpha^2 \big(\sigma_z \rho \sigma_z - 
  \frac{1}{2}\{\sigma_z^2, \rho\}\big),\ \ \ \ \mbox{where} \ \ \ \ 
  \sigma_z = \begin{pmatrix}1 & 0 \\ 0 & -1 \end{pmatrix}.
  \end{equation*} 
In parametrised form equation \eqref{eq Belcom} reads as
    \begin{equation}\label{eq Belcompar}\begin{split}
   d{\bf P}^t = \begin{pmatrix}
  -2\alpha^2 P^t_x - 2B_tP^t_y \\
   -2\alpha^2P^t_y +2B_tP^t_x \\
    0
  \end{pmatrix}dt +
  \begin{pmatrix}
  -2\alpha P^t_y \\
  2 \alpha P^t_x \\
  0
  \end{pmatrix}dW_t.
  \end{split}\end{equation}
     
The dynamics of equation \eqref{eq Belcom} corresponds 
to a two-level atom in a strongly driven, heavily damped,
optical cavity as in \cite{TMW02}, \cite{HSM04}.
The cavity field is assumed to be far off resonance with the 
atomic transition. The cavity is aligned along the $z$-axis 
and instead of controlling with a laser beam as in the previous 
sections, we now control the atom with an external magnetic field 
$B_t$ aligned along the $z$-axis. At the output of the 
cavity we measure the quadrature $Y_t = i(A^*(t)-A(t))$ by a 
homodyne detection scheme. Adiabatic
elimination of the cavity dynamics \cite{TMW02} then 
leads to the dynamics of equation \eqref{eq Belcom}. 
The constant $\alpha$ is determined by properties 
of the cavity and the probe beam \cite{TMW02}.

From the dynamics \eqref{eq Belcom} it follows that $dP^t_z = 0$ and furthermore 
we have
  \begin{equation*}
  d\big({P^t_x}^2+ {P^t_y}^2\big) =
  2P^t_xdP^t_x + dP^t_x dP^t_x + 2P^t_ydP^t_y + dP^t_y dP^t_y = 0,
  \end{equation*}
i.e.\ our problem reduces to a problem on a circle. 
Let us re-parameterize by
introducing $r$ and $\Theta_t$ such that
$P^t_x = r\cos\Theta_t$ and $P^t_y = r\sin\Theta_t$
for $\Theta_t \in [-\pi, \pi)$.
Then the dynamics are given by $dr = 0$ and
  \begin{equation}\label{eq dynpar}
  d\Theta_t = 2B_tdt + 2\alpha dW_t.
  \end{equation}
Replacing $1 -P^T_z$ in the cost functions
\eqref{eq J} and \eqref{eq cost} by
$\Theta_T^2$ will change our goal 
to bringing the system as close as possible to 
the $\sigma_x$-up state at time $t=T$.
It leads to the following expected cost-to-go
function
  \begin{equation*}
  J(t,\, \Theta_t) :=
  \BB{E}_{\Theta_t}\Bigg[\Theta_T^2 + \int_t^T B_s^2  ds\Bigg].
  \end{equation*}
The optimal control problem is now of \emph{linear quadratic}
type, i.e.\ the filtered dynamics are linear and the cost function quadratic. Linear
quadratic problems are well studied and are exactly solvable,
cf.\ \cite{Kus67}, \cite{Kus71}.
 
Starting from \eqref{eq inf}, using It\^o's
formula, we find the following Bellman equation
  \begin{equation*}
  -\partial_tJ^* = \min_{B_t}\Big\{B_t^2+2B_t\partial_\theta J^*\Big\}
  + 2\alpha^2 \partial^2_{\theta\theta}J^*,
  \end{equation*}
with boundary condition $J^*(T,\, \Theta_T) = \Theta_T^2$.
Completing the squares on $B_t$ leads to an optimal
control strategy
  \begin{equation}\label{eq uplusopt}
  B_t = -\partial_\theta J^*,
  \end{equation}
where $J^*$ satisfies the following HJB equation
  \begin{equation}\label{eq HJB}
  -\partial_tJ^*= -{\partial_\theta J^*}^2 + 2\alpha^2\partial^2_{\theta\theta}J^*,
  \end{equation}
with boundary condition $J^*(T, \Theta_T) = \Theta_T^2$.
This equation is solved by making the Ansatz
  \begin{equation*}
  J^*(t,\Theta_t) = \Theta^2_t f(t) + g(t),
  \end{equation*}
for some functions $f$ and $g$.
Substituting this in \eqref{eq HJB} shows that we
have to choose
  \begin{equation*}
  g' = -4\alpha^2 f,
  \end{equation*}
with boundary condition $g(T) =0$.
Furthermore $f$ has to satisfy the \emph{Ricatti equation}
  \begin{equation*}
  f' = 4 f^2,
  \end{equation*}
with boundary condition $f(T) = 1$. Solving these
equations leads to the following expression
for $J^*$
  \begin{equation*}
  J^*(t,\Theta_t) = \frac{\Theta_t^2}{4(T-t)+1} + \alpha^2 \log|4(T-t) +1|,
  \end{equation*}
which satisfies \eqref{eq HJB} as is easily checked.
Equation \eqref{eq uplusopt} now easily leads to
an optimal control strategy given by
  \begin{equation}\label{eq strat}
  B_t = \frac{-2 \Theta_t}{4(T-t) +1}.
  \end{equation}
Summarizing, at time $t$ we have a found measurement 
result $\omega$, integrating the dynamics \eqref{eq dynpar}
we find the state $\Theta_t(\omega)$ and from equation
\eqref{eq strat} we can determine the optimal 
control field $B_t(\omega)$ to be applied at time $t$.

\section{Discussion}

In this paper we have studied the feedback 
control of a qubit in interaction with the 
electromagnetic field. The Belavkin quantum 
filtering equation has been our starting 
point. Introducing the Bloch vector ${\bf P}^t$ as 
a sufficient statistic, as suggested in \cite{Bel83},
we obtained generally non-linear equations 
for the dynamics. In these equations the laser's 
phase and amplitude, represented by $u^+_t$ and $u^-_t$, 
entered as the control parameters. The goal of the control   
was presented by a cost function $J$. 
We proceeded by using the method of dynamic programming 
\cite{Bel57} to find the optimal feedback control strategy. 
In infinitesimal form the dynamic programming algorithm 
leads to the HJB-equation for the optimal cost-to-go function
$J^*$. The optimal control strategy can be expressed in 
terms of the solution to this equation. 

Since the filter equation in general provides non-linear 
dynamics the resulting HJB-equation is often very difficult 
to solve and we have kept this outside the scope of this 
article. Linear dynamics are obtained for systems in which the 
interaction with the environment is \emph{essentially commutative}  
\cite{KuM87}. This means the qubit couples only to one
classical noise of the field. The linear dynamics are obtained
only when the observed process $Y_t$ is exactly this 
classical noise in the field.
In the last section of the article we have 
studied an example of a system for which the dynamics 
are linear. Together with the quadratic cost function 
this lead to an HJB-equation that could be solved 
exactly, providing an explicit expression for 
the optimal control strategy.
 
\section*{Acknowledgments}
\noindent
This work has been sponsored by EC under the
network QP\&Applications (Contract No.\ RTN2-2001-00378).
V.P.B.\ also acknowledges support from EC under the 
program ATESIT (Contract No.\ IST-2000-29681), and 
both S.E.\ and V.P.B. acknowledge support from EPSRC under
the program Mathfit (Grant No. RA2273).

\bibliography{control}

\end{document}